\documentclass[12pt]{JHEP3}         
\usepackage{amssymb,amsfonts}
\usepackage{cite}
\usepackage{amsmath}
\usepackage{graphicx}
\usepackage{MnSymbol}

\def\rmd{{\rm d}}

\newcommand{\el}[1]{\label{#1}}

\newcommand{\nn}{\nonumber \\}

\newcommand{\ba}{\begin{eqnarray}}
\newcommand{\ea}{\end{eqnarray}}
\newcommand{\be}{\begin{equation}}
\newcommand{\ee}{\end{equation}}
\newcommand{\bal}{\begin{align}}
\newcommand{\eal}{\end{align}}
\newcommand{\bay}[1]{\left(\begin{array}{#1}}
\newcommand{\eay}{\end{array}\right)}

\newcommand{\Tr}{\mbox{Tr}}

\def\xG{{\Gamma}}

\def\CN{{\cal N}}

\def\IH{\Bbb{H}}
\preprint{TAU-}
\title{Universal Features of Four-Dimensional Superconformal Field Theory on Conic Space}

\author{Yang Zhou\\ 
School of Physics and Astronomy, Tel-Aviv University\\Ramat-Aviv 69978, Israel\\
{\tt E-mails : yangzhou@post.tau.ac.il}
}

\abstract{Following the set up in arXiv:1408.3393, we study $4d$ $\CN=1$ superconformal field theories on conic spaces. We show that the universal part of supersymmetric R\'enyi entropy $S_q$ across a spherical entangling surface in the limit $q\to 0$ is proportional to a linear combination of central charges, $3c-2a$. This is equivalent to a similar statement about the free energy of SCFTs on conic space or hyperbolic space $\Bbb{S}^1_q\times\Bbb{H}^3$ in the corresponding limit. We first derive the asymptotic formula by the free field computation in the presence of a $U(1)$ R-symmetry background and then provide an independent derivation by studying $\CN=1$ theories on $\Bbb{S}^1_\beta\times\Bbb{S}^3_b$ with a particular scaling $\beta\sim{1\over \sqrt{q}}\,,b=\sqrt{q}$, which thus confirms the validity of the formula for general interacting $\CN=1$ SCFTs. 
Finally we revisit the supersymmetric R\'enyi entropy of general $\CN=2$ SCFTs and find a simple formula for it in terms of central charges $a$ and $c$.
}

\begin{document}

\pagestyle{plain} \setcounter{page}{1}
\newcounter{bean}
\baselineskip16pt


\section{Introduction}
Exact results in interacting quantum field theories (QFTs) are rare in general. The rigid supersymmetry of field theories in curved backgrounds allows one to compute exact results for a certain class of BPS observables~\cite{Nekrasov:2002qd, Pestun:2007rz}. See also~\cite{Kapustin:2009kz, Jafferis:2010un, Hama:2011ea, Hama:2010av, Imamura:2011wg, Alday:2013lba, Hama:2012bg, Nosaka:2013cpa, Closset:2013sxa, Cassani:2014zwa, Assel:2014paa, Kallen:2012cs, Kim:2012ava, Nishioka:2014zpa,Closset:2012ru,Closset:2013vra,Closset:2014uda}. One of such interesting objects is supersymmetric R\'enyi entropy (SRE)~\cite{Nishioka:2013haa}, which in Euclidean signature can be defined using supersymmetric partition function on space with conical singularity (or resolved).

In QFTs in flat space, the R\'enyi entropy can be used to measure the degree of entanglement between two regions $A$ and $\bar A$ separated by the entangling surface $\partial A=\Sigma$. For a state characterized by a density matrix $\rho$ on a spatial slice consisting of $A$ and $\bar A$, one can define the R\'enyi entropy for $A$ using the reduced density matrix $\rho_A=\Tr_{\bar A}\rho$,
\be\label{renyiO}
{S_q={1\over 1-q}\log\Tr[\rho_A^q]\ ,}
\ee where $q$ is a real parameter. The $q\to 1$ limit gives the entanglement entropy (EE) across $\Sigma$. This method to compute entanglement entropy is called replica trick and $q$ is called the replica number. The above formula can be expressed in terms of partition function on a conic space, the $q$-fold cover of the original Euclidean spacetime. The conical singularity generally breaks supersymmetry, therefore the ordinary R\'enyi entropy is non-supersymmetric, however one can maintain the supersymmetry by adding background fields. In consideration of supersymmetry, the SRE is refined to be
\be\label{srenyi}
{S_q={1\over 1-q}\log\left[{Z_q\over Z_1^q}\right]_{SUSY}\ ,}
\ee where the subscript means that all the partition functions appearing in the definition should be supersymmetric. The SRE was first defined and studied in three dimensions~\cite{Nishioka:2013haa, Huang:2014gca, Nishioka:2014mwa} and later generalized to other dimensions~\cite{Huang:2014pda, Crossley:2014oea, Alday:2014fsa, Hama:2014iea}.
In general, these quantities defined by (\ref{srenyi}) are UV divergent but one can extract universal quantities free of ambiguities.

In flat space $\Bbb{R}^{1,d-1}$, the R\'enyi entropy of a $d$-dimensional conformal field theory (CFT) $S_q$ across spherical entangling surface $\Bbb{S}^{d-2}$ can be mapped to that on a sphere with a conical singularity, $\Bbb{S}^d_q$. It can also be mapped to that on a hyperbolic space $\Bbb{S}^1_q\times\IH^{d-1}$, where the entangling surface is mapped to the boundary of $\IH^{d-1}$. We stress that throughout this work, we take the ``smooth cone'' boundary condition~\cite{Lewkowycz:2013laa}, which corresponds to the choice of no boundary contribution in the hyperbolic space.\footnote{See Appendix C in \cite{Lewkowycz:2013laa} for the discussion in details. In particular this choice is consistent with all the known examples of supersymmetric R\'enyi entropy~\cite{Nishioka:2013haa, Huang:2014gca, Nishioka:2014mwa,Huang:2014pda, Crossley:2014oea, Alday:2014fsa, Hama:2014iea}.} The universal part of $S_q$ is shown to be invariant under Weyl transformations of the metric~\cite{Casini:2011kv}
\ba
S_q &=& {1\over 1-q}\left(\log Z_q[\Bbb{S}^d_q] - q\log Z_1[\Bbb{S}^d]\right) \label{renyi1}\\ &=& {1\over 1-q}\left(\log Z_q[\Bbb{S}^1_q\times\IH^{d-1}] - q\log Z_1[\Bbb{S}^1\times\IH^{d-1}]\right)\ .\label{renyi2}
\ea 
This chain of identities are generally true for both non-supersymmetric and supersymmetric R\'enyi entropies of all CFTs. \footnote{Even in the presence of R-symmetry background fields, the relations hold because the Weyl transformation only affects the metric.} 

For $q\to 1$, it is known that the universal part of entanglement entropy is proportional to ``central object'',
\be\label{Seea}
{S_{EE} \propto a_d^*\ ,}
\ee where $a_d^*$ is the a-type central charge in even dimensions and the sphere free energy (over $2\pi$) in odd dimensions. It has also been shown that the R\'enyi entropy near $q=1$ has some universal property~\cite{Perlmutter:2013gua}.\footnote{The specific relation will be modified in the presence of additional background fields.} However, it is not clear whether universal relations exist or not for the R\'enyi entropy at $q$ away from $1$.

In consideration of supersymmetry, the first equation (\ref{renyi1}) enables us to compute the SRE by using the supersymmetric localization technique, while the second equation (\ref{renyi2}) allows us to compute the SRE by using the heat kernel method in the hyperbolic space. Having these computational tools, it is therefore attempting to ask whether one can find universal behaviors of $S_q$ with the help of supersymmetry.

The goal of this work is to show that the universal part of SRE of general $4d$ $\CN=1$ superconformal field theory (SCFT) enjoys a simple relation to the central charges in the limit $q\to 0$, proportional to $3c-2a$, where $a$ and $c$ appear in the Weyl anomaly of the theory~\cite{Duff:1993wm},
\be\label{Weyla}
\langle T_\mu^\mu\rangle \sim {1\over (4\pi)^2} \left(a E-c W^2\right)\ ,
\ee where $W$ is the Weyl tensor and $E$ the Euler density.
More specifically, we will show that the SRE across $\Bbb{S}^2$ in flat space $\Bbb{R}^{1,3}$ for general $4d$ $\CN=1$ SCFTs has the following property at $q\to 0$
\be\label{SREsmallq}
S^{\CN=1}_{q\to 0} = -{4(3c-2a)\over 27q^2}\log(R/\epsilon)\ ,
\ee where $R$ is the radius of the entangling surface $\Bbb{S}^2$ and $\epsilon$ is a short-distance cutoff to regularize the divergence.
This is equivalent to a statement about the free energy of SCFTs on conic space or hyperbolic space $\Bbb{S}^1_q\times\Bbb{H}^3$ in the corresponding limit,
\be\label{freeenergyasm}
F^{\CN=1}_{q\to 0}\propto {4(3c-2a)\over 27}{1\over q^2}\ ,
\ee
which can be considered as the counterpart of the Cardy-like formula on $\Bbb{S}^1\times\Bbb{S}^3$ shown in~\cite{DiPietro:2014bca}, but now for $\Bbb{S}^1\times\Bbb{H}^3$.

We first derive the asymptotic behavior by free field computation in the presence of a $U(1)$ R-symmetry background and then provide a derivation of (\ref{freeenergyasm}) by taking a particular scaling limit of $\CN=1$ partition functions on $\Bbb{S}^1_\beta\times\Bbb{S}^3_b$, which thus shows the universality of the asymptotic behavior for general interacting $\CN=1$ SCFTs. We also revisit the supersymmetric R\'enyi entropy of general $\CN=2$ SCFTs initially studied in~\cite{Huang:2014pda} and find a simple formula for it in terms of central charges $a$ and $c$,
\be
S^{\CN=2} = -\left({c\over q}+4a-c\right)\log(R/\epsilon)\ .
\ee

This paper is organized as follows. In section \ref{asmpRE} we introduce the backgrounds for the discussion of R\'enyi entropy and focus on the the general features in the limits $q\to 0, 1, +\infty$. In particular we extract the universal relations between R\'enyi entropy and other quantities, such as Casimir energy and free energy under those limits. In section \ref{free4dCFTs} we study explicit free field examples. In section \ref{qSCFT1} we construct $\CN=1$ superconformal field theories on the conic sphere $\Bbb{S}^4_q$ and compute free energy as well as supersymmetric R\'enyi entropy. We find that in the limit $q\to 0$, the leading contribution to both of them is proportional to ${4\over 27q^2}(3c-2a)$. We first derive it by free field computation and then derive it by taking use of the exact results from supersymmetric localization. In section \ref{qSCFT2} we revisit $\CN=2$ superconformal field theories on $\Bbb{S}^4_q$ and find simple formulas for the $q$-scalings of free energy as well as supersymmetric R\'enyi entropy in terms of central charges. 

\section{Asymptotic limits of R\'enyi entropy}
\label{asmpRE}
We begin with general discussions on asymptotic behaviors of R\'enyi entropy under different limits, namely $q\to 0, 1, +\infty$. 
\subsection{Backgrounds}
Consider CFTs in a Euclidean flat space $\Bbb{R}^d$ with the metric
\be\label{flatmetric}
ds^2_{\Bbb{R}^d}=d\tau_E^2 + dr^2 + r^2d\Omega_{d-2}^2\ ,
\ee where $\tau_E$ is the Euclidean time and $r\in[0,\infty)$ is the radial direction in the polar coordinates. The entangling surface $\Sigma$ is given by $(\tau_E=0, r=R)$, which is a sub-sphere $\Bbb{S}^{d-2}$. After the transformations
\be\label{transH}
\tau_E= R{\sin\tau\over \cosh \eta + \cos\tau},\quad r= R{\sinh \eta\over \cosh \eta+\cos\tau}\ ,
\ee where $\tau\in[0,2\pi)$ and $\eta\in[0,\infty)$, (\ref{flatmetric}) becomes a hyperbolic space $\Bbb{S}^1\times\Bbb{H}^{d-1}$ with a warp factor $\Omega={1\over \cosh \eta + \cos\tau}$,
\be\label{Hmetric}
ds^2_{\Bbb{R}^d}=\Omega^2ds^2_{\Bbb{S}^1\times\Bbb{H}^{d-1}}=\Omega^2R^2(d\tau^2+d\eta^2+\sinh^2\eta d\Omega^2_{d-2})\ .
\ee
According to equation (\ref{renyi2}), the R\'enyi entropy associated with the spherical entangling surface $\Sigma$ can be computed using the thermal partition function on a hyperbolic space $\Bbb{S}^1_q\times\Bbb{H}^{d-1}$, the metric of which is obtained by $\tau\to q\tau$ for $ds^2_{\Bbb{S}^1\times\Bbb{H}^{d-1}}$ in (\ref{Hmetric}).

According to equation (\ref{renyi1}), one can also compute R\'enyi entropy by working on a branched sphere $\Bbb{S}^d_q$,
\be
ds^2_{\Bbb{S}^d_q}/R^2=\sin^2\theta q^2d\tau^2+d\theta^2+\cos^2\theta d\Omega^2_{d-2}\ ,\label{branmetric}
\ee where $\theta\in [0,\pi/2]$ and $\tau\in[0,2\pi)$, which is conformally equivalent to the hyperbolic space $\Bbb{S}^1_q\times\Bbb{H}^{d-1}$ by the coordinate transformation $\cot\theta=\sinh\eta$
\ba
ds^2_{\Bbb{S}^d_q}=\sin^2\theta R^2(q^2d\tau^2+d\eta^2+\sinh^2\eta d\Omega^2_{d-2}) = \sin^2\theta ds^2_{\Bbb{S}^1_q \times\Bbb{H}^{d-1}}\label{weylmap}\ .
\ea As we see from the above, $\Bbb{S}^d_q$ and $\Bbb{S}^1_q\times\Bbb{H}^{d-1}$ are different by a Weyl factor, which makes that one can compute R\'enyi entropy on either of them. Below we will mostly take the unit $R=1$ and recover it only when it is necessary.

\subsection{Asymptotic limits}
Now we discuss R\'enyi entropy in various limits of the parameter $q$. We are particularly interested in the asymptotic behaviors near $q=0$, $q=1$ and $q=\infty$. Note that R\'enyi entropy $S_q$ has less ambiguity than free energy $F_q:=-\log Z_q$. Namely the R\'enyi entropy is invariant under a shift of free energy by a linear function of $q$. 

In the limit $q\to 1$, one obtains the entanglement entropy in terms of the free energy and its first derivative
\ba\label{ee}
S_{EE}&=& -F_1 + \partial_q F_{q\to 1}\nn &=& -F_1 + 2\pi\partial_\beta F_{\beta\to 2\pi}\ ,
\ea where the length of the Euclidean time circle represents the inverse temperature, $\beta=2\pi q$. It implies that the entanglement entropy is equal to the thermal entropy on a space with a Euclidean time circle, which is conformally equivalent to the original conic space up to a Weyl factor.
The physical meaning of the $\beta$-derivative of $F_\beta$ is the energy
\be\label{energy}
E_q = \partial_\beta F_\beta = {1\over 2\pi} \partial_qF_q\ .
\ee
In the zero temperature limit $\beta\to\infty$, the energy may be defined as the Casimir energy
\be\label{csmr}
E_c := -\lim_{\beta\to\infty}\partial_\beta\log Z = {1\over 2\pi}\lim_{q\to\infty}\partial_qF_q\ ,
\ee which implies that the low temperature expansion of the free energy takes the form
\be\label{lowe}
F_{q\to \infty} = 2\pi q E_c + \cdots\ .
\ee
One can also consider the high temperature expansion ($q\to 0$) of the free energy $F_q$
\be\label{highe}
F_{q\to 0} = F_{(0)} q^{-\alpha} + \cdots\ ,
\ee where in general the free energy does not decrease as temperature goes up, $F_{(0)}\geq0$ and $\alpha\geq 0$. 

{\bf Low temperature expansion}~~~One can substitute (\ref{lowe}) into (\ref{renyi1}) and (\ref{renyi2}) and obtain the asymptotic large $q$ behavior of the R\'enyi entropy
\be\label{asrenyi}
S_{q\to\infty} = 2\pi E_c - F_1\ .
\ee
Note that when $d$ is odd, $\partial_q F_{q\to 1}$ in (\ref{ee}) vanishes since there is no conformal anomaly, therefore one can replace $-F_1$ by $S_{EE}$ and rewrite (\ref{asrenyi}) as
\be\label{largeqlaw}
\Delta S_{q\to\infty}:=S_{q\to\infty}-S_{EE} = 2\pi E_c\ ,
\ee
which may provide a rigorous definition for the Casimir energy $E_c$. 

When $d$ is even and the geometry is branched sphere $\Bbb{S}^d_q$, for the universal logarithmic part, $S_{EE} = -F_1$ is still true and therefore (\ref{largeqlaw}) holds. If the geometry is $\Bbb{S}^1_q\times\IH^{d-1}$ and also $d$ is even, (\ref{largeqlaw}) is no more true because $S_{EE} \neq -F_1$ in this case.\footnote{Note that the condition $\partial_q F_{q\to 1} = 0$ is also true if $F_q$ is refined to be supersymmetric since the ground state energy is zero.}

In general (\ref{largeqlaw}) should be written as
\ba\label{glargeqlaw}
\Delta S_{q\to\infty}&=& 2\pi (E_c - E_1)\nn &:=&2\pi \Delta E\ ,
\ea
which can be obtained by taking the difference between the $q\to\infty$ limit of (\ref{renyi1}) or (\ref{renyi2}) and the equation (\ref{ee}).\footnote{We have inserted (\ref{lowe}).}
This is true generally for all CFTs on a space with a Euclidean time circle, which is conformally equivalent to the original conic space up to a Weyl factor (including both $\Bbb{S}^d_q$ and $\Bbb{S}^1_q\times\IH^{d-1}$).
It means that the relative Casimir energy $E_c - E_1$ can be computed from the zero temperature limit of R\'enyi entropy by the law (\ref{glargeqlaw}).

{\bf High temperature expansion}~~~In the limit $q\to 0$, one can substitute (\ref{highe}) into (\ref{renyi1}) and (\ref{renyi2}) and find that the high temperature behaviors of R\'enyi entropy and free energy coincide (up to a sign flip)
\be\label{smallqlaw}
S_{q\to 0} = -F_{q\to 0}\ .
\ee
Note that so far we have only discussed the leading contributions in both low temperature and high temperature limits. But one can go to the next orders straightforwardly.

{\bf Near ``$q=1$'' expansion}~~~Finally let us look at the behaviors near $q=1$. At $q=1$ we get entanglement entropy according to (\ref{ee}),
\be\label{eeOne}
S_{EE} = -F_1+2\pi E_1\ .
\ee
Unlike $q\to 0$ and $q\to \infty$ cases, we now instead consider the $n$-th $q$-derivative of R\'enyi entropy at $q=1$. One can translate the $q$-derivatives of the R\'enyi entropy into the $q$-derivatives of either free energy or energy~\cite{Perlmutter:2013gua}
\be\label{eeDerivative}
\partial^{(n)}_{q\to 1} S_q = {2\pi\over (n+1)!} \partial^{(n)} _{q\to 1}E_q = {1\over (n+1)!}\partial^{(n+1)} _{q\to 1}F_q\ .
\ee
Furthermore, since the derivative with respect to $q$ can be equivalently considered as the derivative with respect to the metric component $g_{\tau\tau}$, the $n$-th $q$-derivative of R\'enyi entropy at $q=1$ can be shown to be proportional to the integrated $(n+1)$-point function of $T_{\tau\tau}$ in flat space. Indeed the first derivative $\partial_qS_{q\to 1}$ has been shown to be proportional to the coefficient of the stress tensor vacuum two-point function~\cite{Perlmutter:2013gua}. In the presence of supersymmetry, the attention should be paid to the additional R-symmetry background fields.

\section{Free $4d$ CFTs}
\label{free4dCFTs}
In this section we discuss the asymptotic behaviors of R\'enyi entropy by studying free field examples. In particular, we compute (\ref{glargeqlaw})(\ref{smallqlaw})(\ref{eeDerivative}) for the non-supersymmetric R\'enyi entropy of free fields. Notice that the asymptotic coefficients of R\'enyi entropy may be expressed in terms of linear combinations of central charges $a$ and $c$. The linear combinations can be determined by fitting with the explicit results of conformal scalar and massless fermion. However, as we will see that, except for the first $q$-derivative at $q=1$, these relations between non-supersymmetric R\'enyi entropy and central charges are not universal.

We first show the relations coming from experimental test. For $q\to \infty$,
\be\label{largeqac}
\Delta S_{q\to\infty} = {3a+c\over 2}\log(R/\epsilon)\ ,
\ee
while for $q\to 0$,
\be\label{smallqac}
S_{q\to 0} = {a-c\over 2}{1\over q^3}\log(R/\epsilon)+\cdots\ .
\ee
And for the $q$-derivative at $q=1$,
\be\label{qonec}
\partial_qS_{q\to 1}=2c  \log(R/\epsilon)\ .
\ee The first derivative at $q=1$ is consistent with the result in~\cite{Perlmutter:2013gua}.

\subsection{Heat kernel method}
The partition function $Z(\beta)$ on hyperbolic space $\Bbb{S}^1_{\beta=2\pi q}\times\IH^3$ can be computed from the heat kernel of the conformal Laplacian~\cite{Casini:2010kt,Klebanov:2011uf}
\be\label{logZq}
\log Z(\beta) = {1\over 2}\int_0^\infty {dt\over t} K_{\Bbb{S}^1_\beta\times\IH^3}(t)\ ,
\ee
where the kernel of the Laplacian $\Delta$ is defined as $K(t):=\Tr(e^{-\Delta t})$. The kernel on the product space $\Bbb{S}^1_\beta\times\IH^3$ can be factorized
\be\label{Kfactorized}
K_{\Bbb{S}^1_\beta\times\IH^3}(t)= K_{\Bbb{S}^1_\beta}(t) K_{\IH^3}(t)e^{4\pi^2 t}\ ,
\ee
where the exponentiation is to eliminate the gap in the spectrum of the Laplacian on $\IH^3$. The kernel on $\Bbb{S}^1_\beta$ is
\be\label{Kbeta}
K_{\Bbb{S}^1_\beta}(t) = {\beta\over \sqrt{4\pi t}} \sum_{n\neq 0,\in\Bbb{Z}}e^{-{\beta^2n^2\over 4t}+i2\pi n\mu+i\pi n(-1)^f}\ ,
\ee where $f=0$ for scalars and $f=1$ for fermions. 
The hyperbolic space is homogeneous so that the volume can be factorized
\be\label{Khype}
K_{\IH^3}(t) = \int d^3x\sqrt{g} K_{\IH^3}(x,x,t) = V_{\IH^3} K_{\IH^3}(0,t)\ ,
\ee
where $V_{\IH^3}=-2\pi\log(R/\epsilon)$ is the regularized volume of the 3-hyperbolic space $\IH^3$.
$1/\epsilon$ is the IR cut-off and $R$ is the curvature radius of $\IH^3$.

\subsection{Conformal scalar}
For a complex scalar, the equal-point heat kernel on $\IH^3$ is given by
\be\label{Kscalar}
K^s_{\IH^3}(0,t)={2\over (4\pi t)^{3/2}}e^{-4\pi^2t}\ .
\ee
The free energy $F:=-\log Z$ is computed by using (\ref{logZq})(\ref{Kfactorized})(\ref{Kbeta})(\ref{Khype})(\ref{Kscalar})
\be\label{freeenergys}
F^s(\beta) = -{V_{\IH^3}\over 360\pi q^3}\ .
\ee
And the R\'enyi entropy is
\be\label{renyis}
S^s = {(q+1)(q^2+1)V_{\IH^3}\over 360\pi q^3}\ .
\ee
So the limit $q\to \infty$ of the R\'enyi entropy can be extracted
\be\label{lowrenyis}
\Delta S_{q\to\infty} = -{V_{\IH^3}\over 120\pi}\ ,
\ee
and the limit $q\to 0$ of the R\'enyi entropy is
\be\label{highenyis}S_{q\to 0} = {V_{\IH^3}\over 360\pi q^3}\ .
\ee
It can be checked that (\ref{largeqac})(\ref{smallqac}) and (\ref{qonec}) hold, given the central charges $a$ and $c$ for a complex scalar
\be\label{acscalar}a={1\over 180}\ ,\quad c = {1\over 60}\ .
\ee

\subsection{Massless fermion}
For a Weyl fermion, the equal-point heat kernel on $\IH^3$ is
\be\label{Kfermion}
K^f_{\IH^3}(0,t) = {t+2\over (4\pi t)^{3/2}}e^{-4\pi^2t}\ .
\ee
The free energy is then obtained \footnote{Note that there is an additional overall minus sign compared to the scalar.}
\be\label{freeenergyf}
F^f(\beta) = -{(30q^2+7)V_{\IH^3}\over 2880\pi q^3}\ .
\ee
And the R\'enyi entropy is
\be\label{renyif}
S^f = {(q+1)(37q^2+7)V_{\IH^3}\over 2880\pi q^3}\ .
\ee
The limit $q\to \infty$ of the R\'enyi entropy can be extracted
\be\label{lowrenyis}
\Delta S_{q\to\infty} = -{51V_{\IH^3}\over 2880\pi}\ ,
\ee
and the limit $q\to 0$ is
\be\label{highenyis}
S_{q\to 0} = {7V_{\IH^3}\over 2880\pi q^3}\ .
\ee
It can be checked that (\ref{largeqac})(\ref{smallqac}) and (\ref{qonec}) hold, given the central charges $a$ and $c$ for a Weyl fermion
\be\label{acscalar}
a={11\over 720}\ ,\quad c = {18\over 720}\ .
\ee

\subsection{Non-supersymmetric R\'enyi entropy of $\CN=4$ SYM}
It may be a good time to test the universality of (\ref{largeqac})(\ref{smallqac})(\ref{qonec}). Let us now consider the non-supersymmetric R\'enyi entropy of $\CN=4$ Super-Yang-Mills (SYM)~\cite{Fursaev:2012mp}. Free $\CN=4$ Super-Yang-Mills contains $3$ complex scalars, $4$ Weyl fermions and $1$ vector field. Putting all together we have
\be\label{renyiN4SYM}
S_q=3S^s+4S^f+S^v = \frac{\left(1+q+7q^2 +15q^3\right) V_{\IH^3}}{48 \pi q^3}\ ,
\ee
where we have used the R\'enyi entropy of scalar and fermion discussed before and inserted the R\'enyi entropy for a vector field~\cite{Fursaev:2012mp}
\be
S^v = \frac{\left(91 q^3+31 q^2+q+1\right) V_{\IH^3}}{360 \pi  q^3}\ .
\ee

One can check that except for the first $q$-derivative at $q=1$, other relations in (\ref{largeqac})(\ref{smallqac}) are no longer true for this specific example. This may be due to the fact that, the non-supersymmetric R\'enyi entropy of $\CN=4$ SYM is known not protected~\cite{Galante:2013wta}, while central charges $a$ and $c$ are protected. So there is no reason to expect that the asymptotic behaviors of non-supersymmetric R\'enyi entropy of  $\CN=4$ SYM should be characterized by $a$ and $c$. It also leads us to consider the supersymmetric refinement of R\'enyi entropy.

\subsection{Supersymmetric R\'enyi entropy}
Now we start to look at supersymmetric R\'enyi entropy in the free field limit~\cite{Huang:2014pda}. In order to preserve supersymmetry, one has to turn on R-symmetry background fields to twist the boundary conditions.
Turning on a constant background field (chemical potential) along $\Bbb{S}^1_\beta$ gives the kernel $K_{\Bbb{S}^1_\beta}(t)$ a phase shift~\cite{Belin:2013uta}
\be\label{Kbetatilde}
\widetilde K_{\Bbb{S}^1_\beta}(t) = {\beta\over \sqrt{4\pi t}} \sum_{n\neq 0,\in\Bbb{Z}}e^{-{\beta^2n^2\over 4t}+i2\pi n\mu+i\pi n(-1)^f}\ .
\ee
Now one can repeat the free energy computations by using (\ref{logZq})(\ref{Kfactorized})(\ref{Khype})(\ref{Kscalar}). 
For a complex scalar
\be\label{freeenergysA}
F_q^s(\mu) = {\mu^4+2\mu^3+\mu^2-{1\over 30}\over 12\pi q^3}V_{\Bbb{H}^3}\ ,\ee
while for a Weyl fermion,
\be\label{freeenergyfA}
F_q^f(\mu) = {-240\mu^4+120\mu^2+(360\mu^2-30)q^2-7\over 2880\pi q^3}V_{\Bbb{H}^3}\ .\ee
In consideration of supersymmetry, $\mu$ is required to be a function of $q$ and $\mu$ vanishes at $q=1$ because we do not need the R-symmetry background in the absence of conical singularity, $\mu(q=1)=0$. The value of the background field can be solved (by solving Killing spinor equations) either on branched sphere $\Bbb{S}^4_q$ or on hyperbolic space $\Bbb{S}^1_q\times \Bbb{H}^3$ since it is invariant under Weyl transformations.

The supersymmetric R\'enyi entropy can be defined as
\be\label{sre}
S^{SUSY}_q={qF_1(0)-F_q(\mu)\over 1-q}\ .\ee
It is convenient to extract the extra contribution in SRE for each field due to nontrivial $\mu$, $\Delta S=S^{SUSY}_q-S_q$. For a complex scalar
\be\label{sres}
\Delta S^s(\mu) = {\mu^2(\mu+1)^2\over 12\pi(q-1)q^3}V_{\Bbb{H}^3}\ ,\ee
while for a Weyl fermion,
\be\label{sref}
\Delta S^f(\mu) = {\mu^2(-2\mu^2+3q^2+1)\over 24\pi(q-1)q^3}V_{\Bbb{H}^3}\ .\ee
Note that $\Delta S^f(\mu)$ is an even function of $\mu$ but $\Delta S^s(\mu)$ is not. Notice that the effective chemical potential $\mu$ depends on both R-charges of the dynamical fields and the value of the background field.

\section{$\CN=1$ SCFTs}
\label{qSCFT1}
In this section we study $\CN=1$ SCFT with a conserved $U(1)_R$ R-symmetry on four-sphere with a conical singularity, $\Bbb{S}^4_q$. 
As discussed in the introduction, the partition function on $\Bbb{S}^4_q$ can be used to compute supersymmetric R\'enyi entropy. In the absence of the conical deformation, $\CN=1$ SCFTs on round four-sphere were initially discussed in old minimal supergravity~\cite{Festuccia:2011ws} and later discussed in the context of $\CN=1$ conformal supergravity~\cite{Gerchkovitz:2014gta}. These theories can be placed on $\Bbb{S}^4$ canonically since the sphere is conformally flat. It has been shown that the R-current is conserved at the fixed point~\cite{Festuccia:2011ws}.\footnote{We thank Efrat Gerchkovitz and Lorenzo Di Pietro for discussions on this point.} For other systematic study of rigid supersymmetry on four-manifolds, see~\cite{Dumitrescu:2012ha, Dumitrescu:2012at, Klare:2012gn, Cassani:2012ri, Klare:2013dka}.

In the presence of the conical deformation, half of the supersymmetries can remain if one adds a proper R-symmetry background $A_\mu$ to the geometry~\cite{Huang:2014pda}. Once the value of $A_\mu$ is determined, one can perform the heat kernel computation of partition functions by working on $\Bbb{S}^1\times\Bbb{H}^3$ since the Weyl transformation only affects the metric. We will first show that the universal part of supersymmetric R\'enyi entropy $S_q$ (or free energy $F_q$) in $q\to 0$ limit is proportional to a linear combination of central charges $3c-2a$ by free field computation and then provide an independent derivation by taking a particular scaling limit of $\CN=1$ partition functions on $\Bbb{S}^1_\beta\times\Bbb{S}^3_b$, which thus confirms the validity of the asymptotic formula for general interacting $\CN=1$ SCFTs. 
\subsection{Killing spinors on $\Bbb{S}^4_q$}
To construct rigid supersymmetric field theories in curved spacetime, one has to set up Killing spinor equations, whose solutions generate rigid supersymmetries. Those equations will also tell us what background allows rigid supersymmetries. The explicit form of Killing spinors on $\Bbb{S}^4$ depends on the choice of vielbein. In stereographic coordinates, the solutions were well explored in Appendix B in Pestun's work~\cite{Pestun:2007rz}. In spherical coordinates, the solutions were explored first in~\cite{Lu:1998nu}. Now we briefly review $\CN=1$ Killing spinors on conic four-sphere $\Bbb{S}^4_q$~\cite{Huang:2014pda} but leave the details in Appendix \ref{kseS4q}. In 4-spinor notation, the Killing spinor equations take the form of
\ba
& &D_\mu\zeta = +{1\over 2\ell}\gamma_\mu\zeta^\prime\ ,\label{4spinorkilling1}\\
& &D_\mu\zeta^\prime= -{1\over 2\ell}\gamma_\mu\zeta\ ,\label{4spinorkilling2}
\ea where $\zeta$ and $\zeta^\prime$ are Dirac spinors and the covariant derivative is defined as $D_\mu = \nabla_\mu \pm iA_\mu$ to incorporate the R-symmetry background field. 
One can describe the $\Bbb{S}^4_q$ as the one-dimensional blowing up of a $\Bbb{S}^3_q$ or describe it as a $\Bbb{S}^3_q$ fibered on a direction $\rho\in [0,\pi]$~\cite{Huang:2014pda}. We focus on the latter case since the solutions found in the former coordinates~\cite{Huang:2014pda} will project out the $U(1)_R$ symmetry~\cite{Gerchkovitz:2014gta}.
The metric is given by
\be\label{metric22}
d s^2/R^2 = d\rho^2 + \sin\rho^2 (d\theta^2 + \sin^2\theta q^2d \tau^2 + \cos^2\theta d\phi^2)\ .
\ee 
One can check that half of the Killing spinors of the round sphere can be preserved on the $q$-deformed sphere provided the background field is added~\cite{Huang:2014pda}
\be\label{bg}
A_{\Bbb{S}_q^4} = {q-1\over 2}d\tau\ .
\ee See the explicit Killing spinor solutions in Appendix \ref{kseS4q}.

Given the background field (\ref{bg}), now we switch to the hyperbolic space $\Bbb{S}^1_q\times\IH^{3}$ to compute supersymmetric partition function and also supersymmetric R\'enyi entropy.

\subsection{An asymptotic formula at $q\to 0$}
Now we compute supersymmetric partition functions of $\CN=1$ SCFTs on conic space by working on $\Bbb{S}^1_q\times\IH^{3}$ and using heat kernel method.
It is important to emphasize that by working on $\Bbb{S}^1_q\times\IH^{3}$ we only focus on the universal logarithmic term of the corresponding free energy on sphere.

{\bf Free chiral multiplet}~~~A $\CN=1$ chiral multiplet contains 1 complex scalar and 1 Weyl fermion, with R-charge $r$ and $r-1$, respectively. The chemical potentials for the scalar and the fermion are $r(q-1)/2$ and $(r-1)(q-1)/2$, respectively. So the supersymmetric Renyi entropy is given by
\ba\label{srechiral}S^{\text{chiral}} &=& S^s+\Delta S^s\left[{r(q-1)\over 2}\right]+S^f+\Delta S^f\left[{(r-1)(q-1)\over 2}\right]\nn &=& {V_{\Bbb{H}^3}\over 48\pi}\left({2r-3r^2+r^3\over q^2}+{3r^2-2r^3\over q}+2-2r+r^3\right)\ .
\ea

{\bf Free vector multiplet}~~~A $\CN=1$ vector multiplet contains 1 Weyl fermion and 1 vector field, with R-charge $+1$ and $0$ respectively. So the supersymmetric Renyi entropy is given by
\ba\label{srevector}
S^{\text{vector}} &=& S^f + \Delta S^f \left[{q-1\over 2}\right] + S^v\nn &=& {V_{\Bbb{H}^3}\over 24\pi} \left({2\over q}+7\right)\ .
\ea

{\bf Supersymmetric R\'enyi entropy and $a, c$}~~~Consider a $\CN=1$ SCFT which allows a UV free description. $G$ is the gauge group and ${\cal R}_i$ is the representation of the matter field where $i$ denotes different types of matter. One can define the trial functions as
\ba\label{acr}
a&=&{3\over 32}(3\Tr R^3-\Tr R) = {3\over 32}\left[2|G| + \sum_i(3(r_i-1)^3-(r_i-1))|{\cal R}_i|\right],\nn c&=&{1\over 32}(9\Tr R^3-5 \Tr R) = {1\over 32}\left[4|G|+\sum_i(9(r_i-1)^3-5(r_i-1))|{\cal R}_i|\right]\ ,
\ea with $R$ the R-symmetry charge and the trace runs over the fermonic fields of the multiplets of the theory. $|G|$ is the dimension of $G$ and $|{\cal R}_i|$ is the dimension of the representation ${\cal R}_i$.

The SRE for a $\CN=1$ free theory with the gauge group $G$ and matter representation ${\cal R}_i$ is given by
\ba\label{sreSCFT}
S^{\CN=1}_q&=&{V_{\Bbb{H}^3}\over 48\pi}\bigg(\sum_i{2r_i-3r_i^2+r_i^3\over q^2}|{\cal R}_i|+{\sum_i(3r_i^2-2r_i^3)|{\cal R}_i|
+4|G| \over q}\nn &+& \sum_i(2-2r_i+r_i^3)|{\cal R}_i|+14|G|\bigg)\nn
&=&-{4(3c-2a)\over 27}{\log(R/\epsilon)\over q^2}+\cdots ,
\ea
where we have expressed the coefficient of $q^{-2}$ term as a linear combination of functions $a$ and $c$ defined in (\ref{acr}). Therefore we obtain a simple formula for the high temperature limit of $\CN=1$ supersymmetric R\'enyi entropy
\be\label{sresmallq}
S^{\CN=1}_{q\to 0} = -{4(3c-2a)\over 27}{\log(R/\epsilon)\over q^2}\ ,
\ee which can be rewritten in terms of the density
\be\label{densitysre}
s_{q\to 0}:= {S^{\CN=1}_{q\to 0}\over |V_{\Bbb{H}^3}|}=-{2(3c-2a)\over 27\pi}{1\over q^2}\ .
\ee
According to (\ref{smallqlaw}), this can also be translated into the behavior of the density of free energy
\be\label{densityfree}
f_{q\to 0}:={F^{\CN=1}_{q\to 0}\over |V_{\Bbb{H}^3}|}={2(3c-2a)\over 27\pi}{1\over q^2}\ .
\ee

Consider the RG flow from a UV free theory to some IR SCFT, where the $U(1)$ R-symmetry is preserved along the flow. Because of 't Hooft anomaly matching and the algebraic relations between central charges and the $U(1)_R$ anomaly, (\ref{acr}) will determine the IR central charges~\cite{Anselmi:1997am}. The ambiguity of the R-charges $r_i$ of the matter can be fixed by the $a$-maximization~\cite{Intriligator:2003jj}. Therefore at IR fixed point (\ref{sresmallq}) actually characterizes the supersymmetric R\'enyi entropy of the IR SCFT. From the above derivation, it is clear that the formula (\ref{sresmallq}) is true for general $\CN=1$ SCFTs which allow UV free descriptions. Later we will provide an independent derivation of (\ref{densityfree}) from exact results by supersymmetric localization, which thus confirms the universality of the asymptotic behavior for general interacting SCFTs.

{\bf Hofman-Maldacena bound}~~~Now we consider the inequalities satisfied by the R\'enyi entropy derived in the information theory, $H_q:=S_q/S_1$~\cite{Zyczkowski,beckSchlogl}
\ba
\partial_q H_q \leq 0\ ,\label{ineq1}\\
\partial_q\left({q-1\over q}H_q\right)\geq 0\ ,\label{ineq2}\\
\partial_q((q-1)H_q)\geq 0\ ,\label{ineq3}\\
\partial_q^2((q-1)H_q)\leq 0\label{ineq4}\ .
\ea
Imposing these conditions to our result (\ref{sreSCFT}) and take the limit $q\to 0$, one obtains
\be
{a\over c} \leq {3\over 2}\ ,
\ee
which is the Hofman-Maldacena upper bound for general $\CN=1$ SCFTs~\cite{Hofman:2008ar}.

\subsection{Explicit examples}
Now we would like to verify that (\ref{sresmallq}) is true in explicit examples.

{\bf $\CN=1$ description of $\CN=4$ SYM}~~~The simplest $\CN=1$ SCFT may be $\CN=4$ super-Yang-Mills, which can be considered as $\CN=1$ theories with one vector multiplet with gauge group $G$ and 3 chiral superfields $\Phi_{1,2,3}$ in the adjoint representation, coupled through the superpotential $W=h\Phi_1\Phi_2\Phi_3$. One can determine the R-charges for 3 $\CN=1$ chirals ( their R-charges are the same by symmetry ) by the $a$-maximization,
\be\label{amax}
a = {3\over 32}\left[2|G| + 3(3(r-1)^3-(r-1))|G|\right] ,\quad r_* = {2\over 3}\ ,
\ee and the central charges $a,c$ are determined to be 
\be\label{aIR} a=c={|G|\over 4}\ .
\ee Both the gauge coupling and the superpotential can be shown to be exactly marginal using the way in~\cite{Leigh:1995ep}, therefore the central charges $a=c=|G|/4$ are valid for any coupling.

The SRE (\ref{sreSCFT}) is given by
\be\label{sreTrun}
S_q = 3S^{\text{chiral}}(r=2/3) + S^{\text{vector}} = |G|{V_{\Bbb{H}^3}\over 2\pi}{1+7q+19q^2\over 27q^2}\ .
\ee
One can check that
\be\label{Truncheck}
S_{q\to 0} = |G|{V_{\Bbb{H}^3}\over 2\pi}{1\over 27q^2} = {4(2a-3c)\over 27q^2}\log(R/\epsilon)\ ,
\ee
which verifies (\ref{sresmallq}). 

The full SRE result (\ref{sreTrun}) agrees with the SRE obtained by directly dealing with $\CN=4$ SYM with 3 equal $U(1)$ chemical potentials in~\cite{Huang:2014pda}. Furthermore, (\ref{sreTrun}) can be shown to hold at the strong coupling by the holographic computation from $5d$ BPS topological black hole~\cite{Huang:2014pda}, which of course demonstrates the validity of (\ref{Truncheck}) at strong coupling.

Note that one can extract the central charge at UV, $a_{UV}$, from $S_q$ at $q=1$,
\be\label{aUVtrun}
S_{q=1} =  |G|{V_{\Bbb{H}^3}\over 2\pi} := -4a_{UV}\ ,\quad a_{UV}={|G|\over 4}\ ,
\ee which is the same as the central charge $a$ in (\ref{aIR}). This is another way to show that the central charge is independent of marginal couplings in this specific example.

There is another SCFT example including 3 chirals with different R-charges and 1 vector. We leave the analysis in Appendix \ref{3ch1vec}. 

{\bf Klebanov-Witten}~~~Another well-known $\CN=1$ SCFT is the conifold theory proposed by Klebanov and Witten~\cite{Klebanov:1998hh}. It contains 2 vector multiplets and 4 chiral superfields $A_i, B_i (i=1,2)$ in bi-fundamental representations, with a superpotential $W=\lambda \epsilon^{ij}\epsilon^{kl}\Tr(A_iB_kA_jB_l)$. We take the two gauge couplings to be the same, therefore the R-charges of the 4 chirals should be the same by symmetry and one can fix the R-charges by the $a$ maximization,
 \be\label{KWr}
 r_*={1\over 2}\ .
 \ee
Note that the R-symmetry should be non-anomalous. The SRE for this theory is thus given by
\be\label{sreKW}
S_q = |G|(4S^{\text{chiral}}(r=1/2) + 2 S^{\text{vector}}) = |G|{V_{\Bbb{H}^3}\over \pi}{(3+5q(4+13q))\over 96q^2}\ .
\ee
The central charges $a$ and $c$ are given by
\be\label{acKW}
a =c= {3\over 32}|G|\left[2\times 2 + 4(3(r-1)^3-(r-1))/.r\to {1\over 2}\right]={27\over 64}|G|\ .
\ee
Again one can easily verify that
\be S_{q\to 0} = {4(2a-3c)\over 27q^2}\log(R/\epsilon)
\ee
is true. One can also check the famous central charge ratio $32/27$ by comparing (\ref{acKW}) and (\ref{aIR}), given that the central charge of the $\Bbb{Z}_2$ orbifold theory is two times of that of $\CN=4$ SYM.

For the conifold theory, the fix line does not cross the free point, as one can check that
\be\label{seeKW}
S_{q=1} = |G|{V_{\Bbb{H}^3}\over \pi}{88\over 96} = -4{88\over 192}|G|:=-4 a_{UV}\ ,\quad a_{UV} \neq a\ .
\ee
One can extract the ratio between two central charges
\be\label{ratioa}
{a_{UV}\over a} = {88\over 81}\ .
\ee

\subsection{A derivation from exact result}
Now we give a derivation of the high temperature formula of $\CN=1$ partition functions ({\ref{densityfree}) using exact results by supersymmetric localization. The idea is that the hyperbolic space $\Bbb{S}^1_q \times\Bbb{H}^3$ can also be mapped to $\Bbb{S}^1\times\Bbb{S}^3_q$ by a Weyl transformation, where the partition functions can be computed exactly by localization technique in the latter case.

{\bf From hyperbolic space to primary Hopf surface}~~~We first parametrize 3-hyperbolic space $\Bbb{H}^3$ as a fiberation with a 2-torus. The metric of $\Bbb{S}^1_q \times\Bbb{H}^3$ is given by
\be\label{hyperglobal}
ds^2_{\Bbb{S}^1_q \times\Bbb{H}^3}=q^2d\tau^2+\cosh^2\widetilde\eta d\tau_E^2 + d\widetilde\eta^2+\sinh^2\widetilde\eta d\phi^2\ ,
\ee where the domains are given by
\be
\tau\in [0,2\pi)\ ,\quad\tau_E\in [0,2\pi)\ ,\quad\widetilde\eta\in [0,\infty)\ ,\quad \phi\in [0,2\pi)\ . 
\ee This description of $\Bbb{H}^3$ is different from the previous one including a 2-sphere (\ref{Hmetric}). The difference is that here the volume is finite after regularization while the previous volume is logarithmic divergent. Sine $\Bbb{H}^3$ is homogeneous, the free energy on the new description is different from the previous one only by a volume factor. As we will see,
this actually explains why the universal finite part of free energy on a complex manifold could correspond to the universal logarithmic part of the spherical free energy (up to normalization by volume).

By multiplying a factor $\cosh^{-2}\widetilde\eta$ to (\ref{hyperglobal}) and using the coordinate transformation $\sinh\widetilde\eta=\cot\theta$ one obtains
\be
ds^2_{\Bbb{S}^1 \times\Bbb{S}_q^3}=d\tau_E^2 + q^2\sin^2\theta d\tau^2 + d^2\theta+\cos^2\theta d\phi^2\ ,
\ee
which is a direct product of a circle and a 3-dimensional conic sphere, $\Bbb{S}^1\times\Bbb{S}^3_q$.

One can actually define 4-dimensional R\'enyi entropy associated with a 3-dimensional replica number $q$ by using the partition functions on $\Bbb{S}^1\times\Bbb{S}^3_q$.
With supersymmetry, one can instead consider the smooth version with a squashed 3-sphere, $\Bbb{S}^1\times\widetilde{\Bbb{S}}^3_q$, whose metric is given by
\ba\label{Hopfq}
ds^2_{\Bbb{S}^1 \times\widetilde{\Bbb{S}}_q^3}&=&d\tau_E^2 + q^2\sin^2\theta d\tau^2 +f(\theta)^2 d\theta^2+\cos^2\theta d\phi^2\ ,\\
&& f(\theta):=\sqrt{\sin^2\theta+q^2\cos^2\theta}\ .
\ea
This is a primary Hopf surface with a squashed 3-sphere.

It has been shown in~\cite{Huang:2014gca,Alday:2013lba,Closset:2013vra} that 3-dimensional $\CN=2$ supersymmetric partition function only depends on the Reeb vector, which is independent of the resolving factor $f(\theta)$. In 4-dimensional case, it has also been shown in~\cite{Assel:2014paa,Closset:2013vra,Closset:2014uda} that $\CN=1$ supersymmetric partition function only depends on the complex structure, which is independent of the resolving factor too. This is consistent with the 3-dimensional result provided that there is a $4d\to 3d$ reduction. Those results tell us that there is an equivalence between $\CN=1$ partition function on $\Bbb{S}^1\times\widetilde{\Bbb{S}}^3_q$ and that on $\Bbb{S}^1\times\widehat{\Bbb{S}}^3_q$, where $\widehat{\Bbb{S}}^3_q$ is a 3-sphere with a small resolving of the conical singularity, the singular limit of which is used to define the original conic sphere, $\Bbb{S}^3_q$, with ``smooth cone'' boundary condition on it.

{\bf Supersymmetric primary Hopf surface}~~~Now we briefly review the supersymmetric primary Hopf surface with a squashed 3-sphere in the formulation of the rigid limit of new minimal supergravity following~\cite{Dumitrescu:2012ha}.\footnote{See also~\cite{Assel:2015nca,Assel:2014paa}.} In this formulation, the bosonic part of the gravity multiplet includes metric $g_{\mu\nu}$, a R-symmetry gauge field $A_\mu$ and a two form gauge field $B_{\mu\nu}$ whose strength $V^\mu$ is conserved. In Euclidean signature, two independent Killing spinors $\eta$ and $\widetilde\eta$ carrying R-charges $+1$ and $-1$, generally satisfy
\ba\label{ksenew}
(\nabla_\mu -i A_\mu)\,\eta &=& -i V_\mu\eta - iV^\nu \sigma_{\mu\nu}\eta\ ,\nn
(\nabla_\mu +i A_\mu)\,\widetilde\eta &=& +i V_\mu\widetilde\eta + iV^\nu \widetilde\sigma_{\mu\nu}\widetilde\eta\ .
\ea Here $\eta$ and $\widetilde\eta$ are left hand and right hand spinors due to the decomposition of the rotation group $SO(4)=SU(2)_L\times SU(2)_R$. We take the same conventions as that used in~\cite{Dumitrescu:2012ha} in this subsection, see for instance Appendix A there. 

In the coordinates (\ref{Hopfq}), the $U(1)\times U(1)$ isometry of the squashed 3-sphere is generated by the Killing vectors $\partial_\tau$ and $\partial_\phi$ and the third $U(1)$ isometry is generated by $\partial_{\tau_E}$. In the frame,
\be
e^1 = d\tau_{E}\ ,\quad e^2 = f(\theta)d\theta\ ,\quad e^3 = \cos\theta d\phi\ ,\quad e^4 = q\sin\theta d\tau\ ,
\ee
two supercharges $\eta$ and $\widetilde\eta$ carrying opposite R-charges can be obtained by solving (\ref{ksenew})
\be\label{kspinor}
\eta = -{i\over \sqrt{2}}\left(
\begin{array}{c}
e^{{i\over 2}(\phi+\tau-\theta)} \\
ie^{{i\over 2}(\phi+\tau+\theta)} \\
\end{array}
\right)\ ,\quad \widetilde\eta = -{i\over \sqrt{2}}\left(
\begin{array}{c}
e^{-{i\over 2}(\phi+\tau-\theta)} \\
ie^{-{i\over 2}(\phi+\tau+\theta)} \\
\end{array}
\right)\ .
\ee
The complex Killing vector $K:=\eta\sigma^\mu\widetilde\eta\,\partial_\mu$ is given by
\be
K =\partial_{\tau_E} -{i}\partial_\phi -{i\over q}\partial_\tau\ ,
\ee which satisfies
\be
K_\mu K^\mu = 0\ .
\ee
The background fields are given by
\ba
V &=& -{i\over f(\theta)} d\tau_E + \kappa K_\mu dx^\mu\ ,\quad K^\mu\partial_\mu\kappa = 0\ ,\nn
A &=& -{1\over 2 f(\theta)}(2id\tau_E + d\phi + q d\tau) +{1\over 2}(d\phi + d\tau) +{3\over 2}\kappa K_\mu dx^\mu\ .\label{HopfA}
\ea
The contribution (to partition function) of the coupling with the R-symmetry background is controlled by~\cite{Dumitrescu:2012ha}
\be
-iK^\mu A_\mu =-{1\over 2}\left(1+{1\over q}\right)\ . 
\ee
In the limit $q\to 0$, one can find that the dominating component of $A_\mu$ in (\ref{HopfA}) is given by \footnote{The additional minus sign is due to different R-charge conventions on $\Bbb{S}^1\times\Bbb{H}^3$ and $\Bbb{S}^1\times\widetilde{\Bbb{S}}^3_q$.}
\be
-A_\tau = -{1\over 2}\ ,
\ee
which is consistent with the background field (\ref{bg}) on $\Bbb{S}^1\times\Bbb{H}^3$ in the limit $q\to 0$.
Notice that both $\Bbb{S}^1_q\times\Bbb{H}^3$ and $\Bbb{S}^1\times\widetilde{\Bbb{S}}^3_q$ preserve two supercharges, with opposite R-charges.

{\bf Partition function}~~~Before moving on we would like to identify the geometry (\ref{Hopfq}) we are interested in with the familiar one $\Bbb{S}^1_\beta\times\Bbb{S}^3_b$ appearing in the literatures. Recall that the metric of the latter in the usual convention (for instance used in~\cite{Assel:2015nca}) is given by
\be\label{Hopfsq}
ds^2= r_1^2 d\tau_E^2 + r_3^2\left[(b^{-2}\sin^2\theta+b^2\cos^2\theta)d\theta^2+b^{2}\sin^2\theta d\tau^2+b^{-2}\cos^2\theta d\phi^2 \right]\ ,
\ee
where $r_1$ and $r_3$ denote the radius of $\Bbb{S}^1$ and that of the 3-sphere, respectively. $b$ is the squashing parameter for the 3-sphere. (\ref{Hopfq}) and (\ref{Hopfsq}) are the same (up to an overall scale) with the following identifications
\be\label{particularscaling1}
{r_1\over r_3}={1\over \sqrt{q}}\ ,\quad b=\sqrt{q}\ .
\ee
Note that for scale invariant theories the partition function should not depend on the overall scale.
The limit $q\to 0$ corresponds to the zero temperature limit $\beta:=2\pi r_1\to\infty$ with fixed $r_3$. In the squashed 3-sphere's point of view, this is the extremely squashing limit, $b\to 0$.

The partition functions of $\CN=1$ SCFTs on $\Bbb{S}^1_\beta\times\Bbb{S}^3_b$ with the metric (\ref{Hopfsq}) have been studied in~\cite{Assel:2014paa} by supersymmetric localization, see also~\cite{Aharony:2013dha,Assel:2015nca}.\footnote{See~\cite{Romelsberger:2005eg, Kinney:2005ej} for initial works on $\CN=1$ SCFTs on $\Bbb{S}^1\times\Bbb{S}^3$.} In particular, the low temperature contribution to the partition function has been obtained in the leading order
\ba\label{partitioncomplex}
\log Z[\Bbb{S}^1_{\beta\to\infty}\times \Bbb{S}^3_{b=\sqrt{q}}]&=&-\beta E_c + \cdots \nn &\sim& -{r_1\over r_3}\left[{4\pi\over 3}\left({1\over \sqrt{q}}+\sqrt{q}\right)(a-c)+{4\pi\over 27}\left({1\over \sqrt{q}}+\sqrt{q}\right)^3(3c-2a)\right]\ ,\nn
\ea
where $E_c$ is the supersymmetric Casimir energy, which has been shown to be scheme-independent~\cite{Assel:2015nca} and the dots denotes the supersymmetric index part and higher order corrections.
Note that only the ratio ${r_1/r_3}$ appears in the free energy (\ref{partitioncomplex}), which confirms the fact that the final result does not depend on any overall scale of the geometry.

Because of the conformal equivalence between $\Bbb{S}^1_q \times\Bbb{H}^3$ and $\Bbb{S}^1\times\Bbb{S}^3_q$, where the partition function on the latter is equal to that on $\Bbb{S}^1_{\beta=1/\sqrt{q}}\times\Bbb{S}^3_{b=\sqrt{q}}$, together with the fact that the relevant background fields coincide on the two geometries in the limit $q\to 0$, one can actually use the following equality to compute the exact results on the hyperbolic space \footnote{
We also used (\ref{smallqlaw}) in the limit $q\to 0$.}
\be\label{densitybridge1}
f_{q\to 0}[\Bbb{S}^1_q \times\Bbb{H}^3] = f_{q\to 0}[\Bbb{S}^1_{1/\sqrt{q}}\times\Bbb{S}^3_{b=\sqrt{q}}]\ .
\ee
Note that we have normalized the free energy by a $q$-independent volume factor, which in the right hand side of (\ref{densitybridge1}) should be $\text{Vol}(\Bbb{S}^1\times \Bbb{D}^2)=2\pi^2$, according to (\ref{isolatedvol}). In Appendix~\ref{Gcoincidence} we give another geometric reason why the equality (\ref{densitybridge1}) should be true.

Now we apply the scaling (\ref{particularscaling1}) to the free energy result (\ref{partitioncomplex}) and then obtain the free energy in the leading order \footnote{One can check that the supersymmetric index part does not contribute to the leading order in our particular scaling limit. It would be very interesting to establish the precise relation beyond the limit $q\to 0$. We leave this for future work.}
\ba\label{partitionscaling}
f_{q\to 0}[\Bbb{S}^1_q \times\Bbb{H}^3] &=& f_{q\to 0}[\Bbb{S}^1_{1/\sqrt{q}}\times\Bbb{S}^3_{b=\sqrt{q}}]\nn &=& -\lim_{q\to 0}~{\log Z[\Bbb{S}^1_{1/\sqrt{q}}\times\Bbb{S}^3_{b=\sqrt{q}}]\over 2\pi^2}\nn &=& {2\over 27q^2\pi}(3c-2a)\ .
\ea This precisely agrees with the free field result (\ref{densityfree}), which thus confirms the validity of (\ref{densitysre}) and (\ref{densityfree}) for general interacting $\CN=1$ SCFTs.

\section{$\CN=2$ SCFTs revisited}
\label{qSCFT2}
In this section, we revisit $\CN=2$ SCFTs on four-sphere with a conical singularity, $\Bbb{S}^4_q$. The partition function on $\Bbb{S}^4_q$ can be used to compute supersymmetric R\'enyi entropy for $\CN=2$ SCFTs. In the absence of conical deformation, $\CN=2$ SCFTs with Lagrangian descriptions on sphere were extensively studied in~\cite{Pestun:2007rz}, where the spherical partition function and the expectation values of BPS Wilson loop operators were computed by supersymmetric localization. This was extended to squashed four-sphere later in~\cite{Hama:2012bg}. Eventually it was realized that rigid $\CN=2$ supersymmetric field theories on sphere (and deformed) can be well studied in the context of $\CN=2$ conformal supergravity~\cite{Pestun:2014mja}. Recently it was shown that the spherical partition function actually characterizes the Kahler potential~\cite{Gerchkovitz:2014gta}.

In the presence of the conical deformation, it has been shown that part of the supersymmetries can be maintained if one adds a proper $U(1)_J$ R-symmetry background $A^J_\mu$ to the geometry $\Bbb{S}^4_q$~\cite{Huang:2014pda}, where $U(1)_J$ is the diagonal part of $SU(2)_R$ R-symmetry.\footnote{It has been shown in~\cite{Huang:2014pda} that the conic sphere can be defined as the singular limit of a resolved sphere.} The Killing spinor equations have been studied extensively in~\cite{Huang:2014pda} and the value of the background field has been determined
\be
A^J_\tau = {q-1\over 2}\ .
\ee
Given that the background field is invariant under the Weyl transformation between $\Bbb{S}^4_q$ and $\Bbb{S}^1_q\times \Bbb{H}^3$, we can perform the free field computation on the hyperbolic space by heat kernel method once we know the $U(1)_J$ charges for all the dynamical fields.
\subsection{A complete formula}
Now we compute SRE for general $\CN=2$ SCFTs by working on $\Bbb{S}^1_q\times \Bbb{H}^3$ in the free field limit. We focus on the relations between the SRE and central charges $a, c$. 

{\bf Free vector multiplet}~~~A $\CN=2$ vector multiplet includes a $\CN=1$ vector and a $\CN=1$ chiral. In particular we choose the current of the diagonal $U(1)_J$ of $SU(2)_R$ to couple with the background field. Under this $U(1)_J$, the vector component field and the scalar are neutral, and two Weyl fermions are charged oppositely, $\pm 1$~\cite{Seiberg:1994rs}. Note that the supersymmetric R\'enyi entropy for fermions is an even function of the chemical potential. So the total supersymmetric R\'enyi  entropy of a $\CN=2$ vector multiplet is
\be\label{srevvector}
S^V = S^{\text{chiral}}(r=0) + S^{\text{vector}} =  {V_{\Bbb{H}^3}\over 12\pi}\left({1\over q}+4\right)\ .
\ee

{\bf Free hypermultiplet}~~~A $\CN=2$ hypermultiplet contains two $\CN=1$ chiral multiplets, with the scalar and fermion charged $+1$ and $0$ respectively, under $U(1)_J$. So the supersymmetric R\'enyi entropy of a $\CN=2$ hypermultiplet is
\be\label{srehyper}
S^H = 2S^{\text{chiral}}(r=1) = {V_{\Bbb{H}^3}\over 24\pi}\left({1\over q}+1\right)\ .
\ee

{\bf General $\CN=2$ SCFTs}~~~For a general $\CN=2$ SCFT with $n_v$ vectors and $n_f$ hypers, and with $U(1)_J$ being the R-symmetry coupled with the background field, the SRE is given by
\ba\label{sreSSCFT}
S^{\CN=2}_q &=& {V_{\Bbb{H}^3}\over 24\pi}\left({2n_v+n_f\over q} + 8 n_v+n_f\right)\nn &=& -{1\over 12} \left({2n_v+n_f\over q} + 8 n_v+n_f\right)\log(R/\epsilon)\ .
\ea
In particular, for $n_v=n_f$, one obtains
\be\label{sreTWO}
S^{\CN=2}_q = -n_v\left({3q+1\over 4q}\right)\log(R/\epsilon)\ ,
\ee which is true for some specific examples, such as $\CN=2$ description of $\CN=4$ SYM and $\Bbb{S}^5/\Bbb{Z}_2$ orbifold theory. In the former case, (\ref{sreTWO}) agrees with the result obtained in~\cite{Huang:2014pda}. 
 
Recall that the central charge formulae for $\CN=2$ SCFTs which allow weak coupling descriptions
\be\label{SSCFTac}
a={5n_v+n_f\over 24} ,\quad c={2n_v+n_f\over 12}\ .
\ee
One can express the SRE for general $\CN=2$ SCFTs in terms of central charges,
\be\label{sreac}
S^{\CN=2}_q = -\left({c\over q}+4a-c\right)\log(R/\epsilon)\ .
\ee
At $q=1$, (\ref{sreac}) is proportional to $-4a$, which is the standard relation between entanglement entropy and $a$-anomaly. 

Note that the universal log term (\ref{sreSSCFT}) corresponds to the universal logarithmic part of the $\CN=2$ SRE on the $q$-deformed 4-sphere.
Below we will see how to derive (\ref{sreSSCFT}) from the exact $\CN=2$ partition function on $\Bbb{S}^4_q$.

\subsection{A derivation from exact results}
The exact partition function of $\CN=2$ theories on $\Bbb{S}^4_q$ with ``smooth cone'' boundary condition has been obtained in~\cite{Huang:2014pda}
\begin{eqnarray}
 Z =
\int \prod_i \rmd (\sigma)_i\,
 e^{-\frac{8\pi^2}{g_\text{YM}^2}\Tr(\sigma^2)}
 \frac{\prod_{\alpha\in\Delta_+}
 \Upsilon_q( i\sigma\cdot\alpha)
 \Upsilon_q(-i\sigma\cdot\alpha)}
 {\prod_{\cal I}\prod_{\rho\in R_{\cal I}}\Upsilon_q(i\sigma\cdot\rho+\tfrac Q2)}  |Z_\text{inst}|^2\ ,
 \label{Ztot}
\end{eqnarray}
where $\sigma$ is a Lie algebra (of the gauge group) valued constant matrix, $g_\text{YM}$ is Yang-Mills coupling, $\alpha$ is the set of positive roots and $\rho$ is the weight space of the matter, which is in a certain representation of the gauge group. ${\cal I}$ denotes different types of matter. $\Upsilon_q(x)$ is defined to regularize the infinite products as follows
\be
\Upsilon_q(x) = \prod_{m,n\ge0}
 \big(m q^{1/2} +n q^{-1/2}+Q-x\big)
 \big(m q^{1/2} +n q^{-1/2}+x\big)\ ,\quad
Q := \sqrt{q} + 1/\sqrt{q}\ .
\ee
Note that after regularization both the function $\Upsilon_q(x)$ and the matrix integral (\ref{Ztot}) are finite. Nevertheless we want to extract the universal coefficient of the logarithmic divergence on the $q$-deformed 4-sphere from the matrix integral (\ref{Ztot}). 

This can be done by doing large $(\sigma\cdot~)$ expansion for the log of 1-loop determinants and read off the coefficient in front of the log divergence term. This is equivalent to introduce a UV cutoff $\Lambda$ during the (double) Gamma function regularization and pick up the coefficient in front of $\log \Lambda R$~\cite{Kim:2013ola}, which eventually becomes the coefficient of the log divergence term in the final free energy on the $q$-deformed sphere. 

Recall that $\Upsilon_q(x)$ can be decomposed as Barnes double gamma functions
\begin{align}
\Upsilon_q(x)
  = \frac{1}{\xG_2[x, (q^{1/2}, q^{-1/2})]~\xG_2[Q-x,(q^{1/2}, q^{-1/2})]}\ .
\end{align}
At large $|x|$, Barnes double gamma function can be expanded as~\cite{Ruijsenaars:2000}
\ba
\log \xG_2[x,(a,b)] &=& -\frac{1}{2 a b}x^2 \log x + \frac{3}{4 a b} x^2 +\frac{1}{2}  \left(\frac{1}{a}+\frac{1}{b}\right) (x\log x-x)\nn
& &-\left(\frac{1}{12} \left(\frac{a}{b}+\frac{b}{a}\right)+\frac{1}{4}\right)\log x +\cdots\ .\el{G2lartexexp}
\ea
For a pair of vector multiplet, the expansion is the large $x\,(x:=\sigma\cdot\alpha)$ expansion for $\log[\Upsilon_q(ix)\Upsilon_q(-ix)]$, which is given by
\ba\label{vectorexpansion}
&&\log[\Upsilon_q(ix)\Upsilon_q(-ix)]\nn &&= 3x^2+\left(-2x^2 + {1+3q+q^2\over 3q}\right)\log x +{(1+q)^2\over 12q x^2} + {\cal O}(x^{-4})\ .
\ea
For a pair of hypermultiplet, the expansion is the large $y\,(y:=\sigma\cdot\rho)$ expansion for $-2\log[\Upsilon_q(iy+Q/2)]$, which is given by
\ba\label{hyperexpansion}
&&-2\log[\Upsilon_q(iy+Q/2)]\nn &&= -3y^2+\left(2y^2 + {1+q^2\over 6q}\right)\log y - {(1+q)^2(1+6q+q^2)\over 96q^2y^2} + {\cal O}(y^{-4})\ .
\ea
The quadratic terms in (\ref{vectorexpansion}) and (\ref{hyperexpansion}) cancel with each other because of the conformality condition~\cite{Pestun:2007rz}
\be
\sum_\alpha (\sigma\cdot\alpha)^2 = \sum_\rho (\sigma\cdot\rho)^2\ .
\ee
One can add up the log coefficient of (\ref{vectorexpansion}) and the log coefficient of (\ref{hyperexpansion}) because both of them can be translated into the coefficient of $\log\Lambda R$ as mentioned before. 

For a general $\CN=2$ conformal field theory containing $n_v$ vector multiplets and $n_f$ hypermultiplets, one can again take use of the conformality condition and obtains
the $q$ scaling of the universal term of free energy 
\be\label{finalfreeenergy}
F^{\CN=2}_q \propto\left( {n_v\over 2}{1+3q+q^2\over 3q} + {n_f\over 2}{1+q^2\over 6q}\right)\ ,
\ee
which gives the $q$-scaling of the supersymmetric R\'enyi entropy
\be\label{finalsre}
S^{\CN=2}_q \propto -\left({4q+1\over 6q} n_v + {q+1\over 12q}n_f\right)=-\left({c\over q}+4a -c\right)\ .
\ee This precisely agrees with the free field result (\ref{sreSSCFT}), which implies that the results (\ref{sreSSCFT}) and (\ref{sreac}) are universal. It would be interesting if one can reproduce the $q$-scalings from the background invariants in the context of $\CN=2$ conformal supergravity.

Note that the large $(\sigma\cdot~)$ expansion we used here is also consistent with the large mass expansion in the works~\cite{Hertzberg:2010uv,Lewkowycz:2012qr}, which turns out to be the correct way to extract universal entanglement entropy (and R\'enyi entropy) for massive theories. Recall that in the weak coupling limit $g_{YM}\to 0$, the 1-loop part in (\ref{Ztot}) is essentially the effective action with a background field $\sigma$ on the $q$-deformed sphere (which can be used to compute R\'enyi entropy in free field limit). Now consider the numerator and denominator of the 1-loop part in (\ref{Ztot}) separately, the effective actions are for massive theories, since the coupling with $\sigma$ gives the effective mass term both for the vector multiplet and the hypermultiplet. The universal parts of R\'enyi entropy (supersymmetric) for these massive theories come from the log coefficient in (\ref{vectorexpansion}) and that in (\ref{hyperexpansion}), respectively, following~\cite{Hertzberg:2010uv,Lewkowycz:2012qr}. Further cancellation because of the conformality condition eliminates the mass dependence, which eventually leads to the final results (\ref{finalsre}).

{\bf Bounds of $a/c$}~~~Now we consider the inequalities for $\CN=2$ supersymmetric R\'enyi entropy (\ref{sreac})
\be
H_q:={S_q\over S_1} = {{c\over q}+4a-c\over 4a}\ .
\ee
The large $q$ limit of (\ref{ineq2}) gives
\be
{a\over c}\geq {1\over 2}\ ,
\ee which is the lower bound for $\CN=2$ theories consistent with the results in~\cite{Shapere:2008zf,Hofman:2008ar}.

\section*{Acknowledgement}
We are grateful for helpful discussions with Ofer Aharony, Lorenzo Di Pietro, Efrat Gerchkovitz, Xing Huang, Zohar Komargodski, Yaron Oz, Soo Jong Rey, Amit Sever, Cobi Sonnenschein. This work was supported by ``The PBC program for fellowships for outstanding post-doctoral researcher from China and India of the Israel council of higher education''.

\appendix
\section{Killing spinors on $\Bbb{S}^4_q$}\label{kseS4q}
The round four-sphere can be considered as a three-sphere fibered on the $\rho$ direction. The metric is given by
\be\label{metric22}
\rmd s^2/R^2 = \rmd\rho^2 + \sin\rho^2 (\rmd\theta^2 + \sin^2\theta\rmd \tau^2 + \cos^2\theta\rmd\phi^2)\ ,
\ee with
\be
\rho\in[0, \pi]\ ,\quad \theta\in[0,\pi/2]\ ,\quad \tau\in[0,2\pi)\ ,\quad \phi\in[0,2\pi)\ .
\ee
 We choose the vielbein as
\ba
e^1/R &=& \sin\rho \sin(\tau+\phi) \rmd\theta + \sin\rho \cos(\tau + \phi) \sin\theta \cos\theta (\rmd\tau -\rmd\phi)\ ,\nn
e^2/R &=& -\sin\rho \cos(\tau+\phi) \rmd\theta + \sin\rho \sin(\tau + \phi) \sin\theta \cos\theta (\rmd\tau -\rmd\phi)\ , \nn
e^3/R &=& \sin\rho\,(\sin\theta^2\rmd\tau + \cos\theta^2\rmd\phi)\ ,\quad e^4/R = \rmd \rho\ .\label{vielbein22}
\ea
One can define a $T$ matrix as 
\be
T(\rho) = \left(
\begin{array}{cccc}
 0 & 0 & -\frac{1}{2} \tan \frac{\rho }{2} & 0 \\
 0 & 0 & 0 & -\frac{1}{2} \tan \frac{\rho }{2} \\
 \frac{1}{2} \cot \frac{\rho }{2} & 0 & 0 & 0 \\
 0 & \frac{1}{2} \cot \frac{\rho }{2} & 0 & 0 \\
\end{array}
\right)\ .
\ee We take the gamma matrices given in terms of Pauli matrices as follows
\ba\label{gammas}
\gamma_1= \left(
\begin{array}{cc}
 0 & i\tau_1 \\
 -i\tau_1 & 0 \\
\end{array}
\right)\ ,\quad \gamma_2=\left(
\begin{array}{cc}
 0 & i\tau_2 \\
 -i\tau_2 & 0 \\
\end{array}
\right)\ ,\nn
\gamma_3=\left(
\begin{array}{cc}
 0 & i\tau_3 \\
 -i\tau_3 & 0 \\
\end{array}
\right)\ ,\quad \gamma_4=\left(
\begin{array}{cc}
 0 & 1_{2\times 2} \\
  1_{2\times 2}  & 0 \\
\end{array}
\right)\ .
\ea
And the chiral projection matrices are given by
\be
P_L = {1\over 2}(1+\gamma_5)\ ,\quad P_R = {1\over 2}(1-\gamma_5)\ ,
\ee where
\be
\gamma_5=\gamma_1\gamma_2\gamma_3\gamma_4\ .
\ee
With the vielbein and metric given above, one can actually find the following
matrix identities
\be
{1\over 4}\omega_\mu = \gamma_\mu T(\rho)\ ,\quad \mu = \theta, \tau, \phi
\ee where $\omega_\mu$ are spin connections. It means that an arbitrary constant 4-spinor $\zeta_0$ satisfies the first three equations ($\mu=\theta, \tau, \phi$) of (\ref{4spinorkilling1}), provided that $\zeta^\prime$ is defined as
\be\label{4spinorkilling11}
{1\over 2\ell}\zeta^\prime := T(\rho) \zeta\ . 
\ee
There remains an undetermined $\rho$-dependent matrix factor $S(\rho)$ and the true Killing spinor will be given by
\be
\zeta = S(\rho) \zeta_0\ .
\ee 
Finally, $S(\rho)$ can be determined by studying the $\rho$ component of equation (\ref{4spinorkilling1}) and it is given by
\be
S(\rho)=\left(
\begin{array}{cccc}
 \sin \frac{\rho }{2} & 0 & 0 & 0 \\
 0 & \sin \frac{\rho }{2} & 0 & 0 \\
 0 & 0 & \cos \frac{\rho }{2} & 0 \\
 0 & 0 & 0 & \cos \frac{\rho }{2} \\
\end{array}
\right)\ .
\ee
Now the $q$-branched four-sphere is obtained by replacing $\rmd\tau$ in (\ref{metric22})(\ref{vielbein22}) by $q\rmd\tau$, and it is easy to check that 
\be\label{zeta3}
\zeta = S(\rho)\left(
\begin{array}{c}
c_1 \\
c_2 \\
c_3 \\
c_4 \\
\end{array}
\right)
\ee
is still a solution, provided that a background field is turned on through the coupling
\be
\widetilde D_\mu=\nabla_\mu + i A_\mu \left(
\begin{array}{cccc}
 +1 & 0 & 0 & 0 \\
 0 & -1 & 0 & 0 \\
 0 & 0 & +1 & 0 \\
 0 & 0 & 0 & -1 \\
\end{array}
\right)\ ,
\ee where $A_\mu$ takes the value
\be
A_{\mathbb{S}^4_q} = {q-1\over 2}\rmd\tau\ .
\ee
Recall that $P_L$ and $P_R$ project out the upper half and lower half of the 4-spinors. $P_L\zeta$ and $P_R\zeta$ have opposite R-charges, $+1$ and $-1$. Therefore the true coupling with the background field is
\be
D_\mu=\nabla_\mu + i A_\mu \left(
\begin{array}{cccc}
 +1 & 0 & 0 & 0 \\
 0 & +1 & 0 & 0 \\
 0 & 0 & -1 & 0 \\
 0 & 0 & 0 & -1 \\
\end{array}
\right)\ .
\ee
This means that the second component and the third component of the Killing spinor were killed out and the remaining ones have opposite R-charges,
\be\label{zeta4}
\zeta = S(\rho)\left(
\begin{array}{c}
c_1 \\
0 \\
0 \\
c_4 \\
\end{array}
\right)\ .
\ee

\section{$3$ different chirals plus $1$ vector}
\label{3ch1vec}
Now we consider SCFTs including $3$ chirals with different R-charges $r_1,r_2,r_3$ and $1$ vector, all standing in the adjoint representation of some gauge group $G$. They may be realized as IR fixed points of some RG flows with UV free descriptions and nontrivial potentials.
Let us parametrize the three R-charges as 
\be r_1={2x\over 3}\ , r_2={2y\over 3}\ , r_3={2z\over 3}\ .
\ee The SRE is given by
\ba\label{srethreeR}
S_{q}= |G|{V_{\Bbb{H}^3}\over 2\pi} &&{A+(3B-2A)q+(27-3B+A)q^2\over 27q^2}\ ,\nn
&&A=xyz ,\quad B=xy+yz+zx\ ,
\ea
where the sum of R-charges satisfies $x+y+z=3$ by the constraint from the cubic superpotential. One specific way to construct this IR SCFT is to consider $\CN=4$ SYM under 3 different $U(1)$ chemical potentials. And (\ref{srethreeR}) precisely agrees with the SRE computed by directly dealing with $\CN=4$ SYM in~\cite{Huang:2014pda}. The result (\ref{srethreeR}) has also been shown~\cite{Huang:2014pda} to hold in the strong coupling by the holographic computation of SRE from $5d$ BPS STU topological black hole~\cite{Behrndt:1998ns, Behrndt:1998jd, Cvetic:1999xp} with three different charges.

Let us look at the central charges of the IR SCFT. $a$ and $c$ are given by
\be\label{threeRac}
a=c= |G|{9\over 32}(1+\sum_i^3 (r_i-1)^3)={|G|\over 4}xyz\ .
\ee
One can verify that (\ref{sresmallq}) is again satisfied.

By looking at the $q\to 0$ limit of (\ref{srethreeR}) we clearly see that the R-charges affect the high temperature behavior of SRE, therefore affect the central charges. However this does not happen when $q\to 1$, where
\be
S_{q\to 1}=|G|{V_{\Bbb{H}^3}\over 2\pi}\ ,\quad a_{UV}={|G|\over 4}\ ,
\ee which is independent of R-charges.

\section{A geometric coincidence}
\label{Gcoincidence}
Let us first look into more details about the scaling geometry under the limit $q\to 0$ for $\Bbb{S}^1_q\times\Bbb{H}^3$. The metric of $\Bbb{S}^1_q\times\Bbb{H}^3$ is given by
\be
ds^2_{\Bbb{S}^1_q \times\Bbb{H}^3}=R^2(q^2d\tau^2+d\eta^2+\sinh^2\eta d\Omega^2_2)\ ,
\ee
which can be related to the $\tau$ circle fibered on a 3-dimensional flat space by a Weyl factor $f:=\cosh^{-2}\eta$, 
\be\label{weylfiber}
ds^2_{\Bbb{S}^1_q \times\Bbb{H}^3}=R^2f^{-1}(f q^2d\tau^2+f (d\eta^2+\sinh^2\eta d\Omega^2_2))\ ,
\ee where $f$ is chosen to make the metric $f ds^2_{\Bbb{H}^3}$ flat,
\be
f ds^2_{\Bbb{H}^3} = dr^2 + r^2 d\Omega_2^2\ ,\quad r=\tanh\eta\in[0,1]\ .
\ee
 After dropping the overall factor $R^2f^{-1}$, the metric (\ref{weylfiber}) becomes a circle fibered on a flat 3-space, 
 \be
 ds^2 = (1-r^2)q^2 d\tau^2 + dr^2 + r^2d\Omega_2^2\ , \quad r\in [0,1]\ ,
 \ee
 where the universal part of R\'enyi entropy or supersymmetric R\'enyi entropy should be the same as that on the hyperbolic space since they are conformally equivalent~\cite{Casini:2011kv}. Under the limit $q\to 0$, the $\tau$ circle shrinks and the geometry goes to a product with a tiny circle
\be\label{scalingspace}
\Bbb{S}^1_{q\to 0}\times\Bbb{R}^3\ .
\ee Notice that the supersymmetric nature is preserved on the space (\ref{scalingspace}), which can be seen by noting that the heat kernel on $\Bbb{S}^1_q$ (\ref{Kbetatilde}) has periodic boundary condition for fermions in the limit $q\to 0$. 

The key observation is that the space (\ref{scalingspace}) can also be obtained by taking a particular limit of $\Bbb{S}^1_\beta\times\Bbb{S}^3_b$. This is the limit
\be\label{particularlimit}
\beta\sim {1\over \sqrt{q}}\to {1\over 0}\ ,\quad b=\sqrt{q}\to 0\ .
\ee 

Let us look into more details about the scaling limit of $\Bbb{S}^1_\beta\times\Bbb{S}^3_b$. We use $r_1$ and $r_3$ to denote the radius of the circle and the 3-sphere, respectively. The length of the circle is defined as the inverse temperature, $\beta=2\pi r_1$. The metric of $\Bbb{S}^1_\beta\times \Bbb{S}^3_b$ is given by
\be\label{hopfmetric}
ds^2= r_1^2 d\tau^2 + r_3^2\left[(b^{-2}\sin^2\theta+b^2\cos^2\theta)d\theta^2+b^{-2}\cos^2\theta d\phi_1^2 +b^{2}\sin^2\theta d\phi_2^2\right]\ ,
\ee where
\be
\tau\in [0,2\pi)\ ,\quad \theta\in [0,{\pi/ 2}]\ ,\quad \phi_1\in [0,2\pi)\ ,\quad \phi_2\in [0,2\pi)\ .
\ee Replacing $b$ by $\sqrt{q}$ one obtains
\be\label{hopfmetricc}
{ds^2\over r^2_3} = {r_1^2\over r_3^2} d\tau^2 + {1\over q}\left[(\sin^2\theta+q^2\cos^2\theta)d\theta^2+\cos^2\theta d\phi_1^2 +q^2\sin^2\theta d\phi_2^2\right]\ ,
\ee where we have divided the metric by an overall $r_3$, which will not affect the final partition function for scale invariant theories. Notice that the part inside the square brackets is the 
squashed 3-sphere $\widetilde{\Bbb{S}}^3_q$, which can be considered as the resolved version of (\ref{branmetric}) in $d=3$. 

Under the limit $q\to 0$, 
\ba
ds^2_{\widetilde{\Bbb{S}}^3_q} &=& \sin^2\theta d\theta^2+\cos^2\theta d\phi_1^2 +q^2\sin^2\theta d\phi_2^2\nn
&=& dy^2 + y^2d\phi_1^2 +q^2(1-y^2) d\phi_2^2\ ,\quad (y:=\cos\theta\in [0,1])
\ea which is effectively $\Bbb{D}^2\times \Bbb{S}^1_q$ in the limit. Here $\Bbb{D}^2$ means 2-disk with unit radius. In consideration of the $1/q$ factor in front of the square brackets in (\ref{hopfmetricc}), the geometry becomes $\Bbb{D}^2_{1/\sqrt{q}}\times \Bbb{S}^1_{\sqrt{q}}$.

Now we take a particular scaling for $r_1$ in order to combine the circle $\Bbb{S}^1_\beta$ and the 2-disk $\Bbb{D}^2$ we have just obtained from the extremely squashed 3-sphere,
\be\label{particularscaling}
{r_1^2\over r_3^2} = {1\over q}\ .
\ee
Putting all together in the limit $q\to 0$, the 4-geometry becomes
\be
[\Bbb{S}^1\times \Bbb{D}^2]_{1/\sqrt{q}}\times \Bbb{S}^1_{\sqrt{q}}\ ,
\ee which can be rescaled by multiplying an overall factor $\sqrt{q}$,
\be\label{isolatedvol}
[\Bbb{S}^1\times \Bbb{D}^2] \times \Bbb{S}^1_q\ .
\ee In the limit $q\to 0$, the geometry becomes
\be
\Bbb{S}^1_{q\to 0}\times\Bbb{R}^3\ ,
\ee which is the same as the one in (\ref{scalingspace}). 

Because of the coincidence of different geometries in the limit $q\to 0$, we could be able to compute the exact result on $\Bbb{S}^1_q \times\Bbb{H}^3$ using the available results on other manifold, such as $\Bbb{S}^1_\beta\times\Bbb{S}^3_b$, by identifying the leading free energy (normalized by a volume factor)
\be\label{densitybridge}
f_{q\to 0}[\Bbb{S}^1_q \times\Bbb{H}^3] = f_{q\to 0}[\Bbb{S}^1_{1/\sqrt{q}}\times\Bbb{S}^3_{b=\sqrt{q}}]\ .
\ee

\end{document}